\documentclass[sigconf]{acmart}
\usepackage{graphicx}
\usepackage{subcaption}
\usepackage{enumitem}
\usepackage{pifont}
\usepackage[skip=0pt]{caption} 
\usepackage{titlesec}
\titlespacing{\section}{0pt}{*0.5}{*0.5} 
\titlespacing{\subsection}{0pt}{*0.4}{*0.4} 
\AtBeginDocument{%
  }

\copyrightyear{2025}
\acmYear{2025}
\setcopyright{acmlicensed}\acmConference[WWW Companion '25]{Companion Proceedings of the ACM Web Conference 2025}{April 28-May 2, 2025}{Sydney, NSW, Australia}
\acmBooktitle{Companion Proceedings of the ACM Web Conference 2025 (WWW Companion '25), April 28-May 2, 2025, Sydney, NSW, Australia}
\acmDOI{10.1145/3701716.3715199}
\acmISBN{979-8-4007-1331-6/25/04}

\begin{document}
\title{SortingHat: Redefining Operating Systems Education with a Tailored Digital Teaching Assistant}
\author{Yifan Zhang}
\orcid{0009-0001-1416-2073}
\affiliation{%
  \institution{School of Software Technology, Zhejiang University}
  \city{Ningbo}
  \state{Zhejiang}
  \country{China}
}
\email{yifanzhanag@zju.edu.cn}

\author{Xinkui Zhao\textsuperscript{*}}
\orcid{0000-0002-1115-5652}
\affiliation{%
  \institution{School of Software Technology, Zhejiang University}
  \city{Ningbo}
  \state{Zhejiang}
  \country{China}
}
\email{zhaoxinkui@zju.edu.cn}

\author{Zuxin Wang}
\orcid{0009-0004-9558-5490}
\affiliation{%
  \institution{School of Software Technology, Zhejiang University}
  \city{Ningbo}
  \state{Zhejiang}
  \country{China}
}
\email{wangzuxin@zju.edu.cn}

\author{Zhengyi Zhou}
\orcid{0009-0009-5239-6151}
\affiliation{%
  \institution{School of Software Technology, Zhejiang University}
  \city{Ningbo}
  \state{Zhejiang}
  \country{China}
}
\email{zhouzhengyi@zju.edu.cn}

\author{Guanjie Chen}
\orcid{0000-0003-2080-3903}
\affiliation{%
  \institution{School of Software Technology, Zhejiang University}
  \city{Ningbo}
  \state{Zhejiang}
  \country{China}
}
\email{chengguanjie@zju.edu.cn}

\author{Shuiguang Deng}
\orcid{0000-0001-5015-6095}
\affiliation{%
  \institution{School of Software Technology, Zhejiang University}
  \city{Ningbo}
  \state{Zhejiang}
  \country{China}
}
\email{dengsg@zju.edu.cn}

\author{Jianwei Yin}
\orcid{0000-0003-4703-7348}
\affiliation{%
  \institution{School of Software Technology, Zhejiang University}
  \city{Ningbo}
  \state{Zhejiang}
  \country{China}
}
\email{zjuyjw@cs.zju.edu.cn}

\vspace{-10pt} 
\vspace{-10pt} 

\renewcommand{\shortauthors}{Yifan Zhang et al.}
\begin{abstract}
Operating Systems (OS) courses are among the most challenging in computer science education due to the complexity of internal structures and the diversity of running environments. Traditional teaching methods often fail to address the diverse backgrounds, learning speeds, and practical needs of students. To tackle these challenges, we present SortingHat, a personalized digital teaching assistant tailored specifically for OS education. SortingHat integrates advanced AI technologies, including a retrieval-augmented generation (RAG) framework and multi-agent reinforcement learning (MARL), to deliver adaptive, scalable, and effective educational support. SortingHat features a 3D digital human interface powered by large language models (LLMs) to provide personalized, empathetic, and context-aware guidance. It generates tailored exercises based on each student’s learning history and academic performance, reinforcing weak areas and challenging advanced concepts. Additionally, the system incorporates a robust evaluation pipeline that ensures fair, consistent, and unbiased grading of student submissions while delivering personalized, actionable feedback for improvement. By combining personalized guidance, adaptive content creation, and automated assessment, SortingHat transforms OS education into an engaging, immersive, and scalable experience.
\end{abstract}

\ccsdesc[300]{Information systems~Users and interactive retrieval}
\ccsdesc[500]{Computing methodologies~Knowledge representation and reasoning}
\ccsdesc[500]{Human-centered computing~Interaction design}
\vspace{-10pt} 
\keywords{Education, Large Language models, Digital Human, Retrieval Augmented Generation,Multi Agent Reinforcement Learning}
\vspace{-10pt} 

\maketitle

\vspace{-10pt}

\section{Introduction}
Operating Systems (OSs) courses are widely recognized as some of the most challenging subjects in computer science education. The inherent complexity of OS concepts—ranging from process management and memory allocation to kernel development—stems not only from the need to master both theoretical principles and practical applications but also from the diversity of hardware architectures and the wide range of application environments. Additionally, students often have diverse educational backgrounds and varying levels of foundational knowledge, making it difficult for traditional teaching methods to meet their individual needs effectively. Learning OS concepts is a gradual and iterative process, yet a one-size-fits-all approach fails to accommodate students’ unique learning paces and interests. Moreover, OS courses are highly practice-oriented, emphasizing hands-on experimentation with real-world systems. From designing kernel modules to debugging system calls, students must apply theoretical knowledge to practical scenarios, which often require immediate, context-sensitive guidance. Unfortunately, traditional teaching resources, such as static lecture materials or limited office hours, are insufficient to address these challenges, leaving many students struggling to keep up. Even outside the classroom, students working on OS-related projects or research face similar difficulties, further underscoring the need for scalable, adaptive support. 

To address these issues, we present SortingHat, a personalized education system designed specifically to enhance OS learning. SortingHat leverages advanced AI technologies to tackle the unique challenges of OS education through three key functionalities. First, SortingHat features a customizable digital human teaching assistant powered by large language models (LLMs). This assistant not only provides expert-level guidance on OS concepts but also adapts its responses based on each student’s learning progress. By delivering answers in an engaging and context-aware manner, it ensures personalized and empathetic support that fosters deeper understanding and motivation. Second, SortingHat utilizes LLMs to generate customized exercises tailored to students’ individual learning histories, performance, and feedback. These exercises focus on reinforcing weak areas and challenging advanced concepts, ensuring that learning remains targeted, adaptive, and effective. Finally, SortingHat incorporates a multi-agent reinforcement learning (MARL) evaluation system that provides consistent, unbiased grading of student submissions. This system ensures fairness and reliability while offering personalized feedback and learning suggestions, helping students address specific weaknesses and improve continuously.

By integrating these three core functionalities—personalized guidance, adaptive exercises, and fair assessment—SortingHat offers a comprehensive solution to the challenges of OS education. It not only empowers students with tailored learning experiences but also reduces the workload for educators, making it a scalable and effective tool for modern education.

Our contributions are:
\begin{itemize}[topsep=0pt, partopsep=0pt, itemsep=1pt, parsep=0pt]
    \item SortingHat features a lifelike 3D digital human assistant powered by LLMs. This assistant provides expert-level, context-aware guidance on OS concepts and adapts its responses to students' learning progress. By combining technical expertise with emotional interaction, it enhances engagement and creates a supportive learning environment.
    \item Leveraging LLMs, SortingHat generates customized exercises tailored to individual students’ learning histories, performance, and feedback. These exercises are designed to address weak areas, reinforce foundational knowledge, and challenge advanced concepts, ensuring a personalized and targeted learning experience that adapts to each student’s unique needs.
    \item  SortingHat introduces a MARL evaluation system to ensure accurate, unbiased, and consistent grading of student submissions. This system also provides personalized feedback and actionable learning suggestions, enabling students to address their weaknesses and improve iteratively.
\end{itemize}
\vspace{-10pt} 

\section{SortingHat Overview}
    \vspace{-5pt} 

\begin{figure}[h]
    \vspace{-5pt} 
    \centering
    \includegraphics[width=1\linewidth]{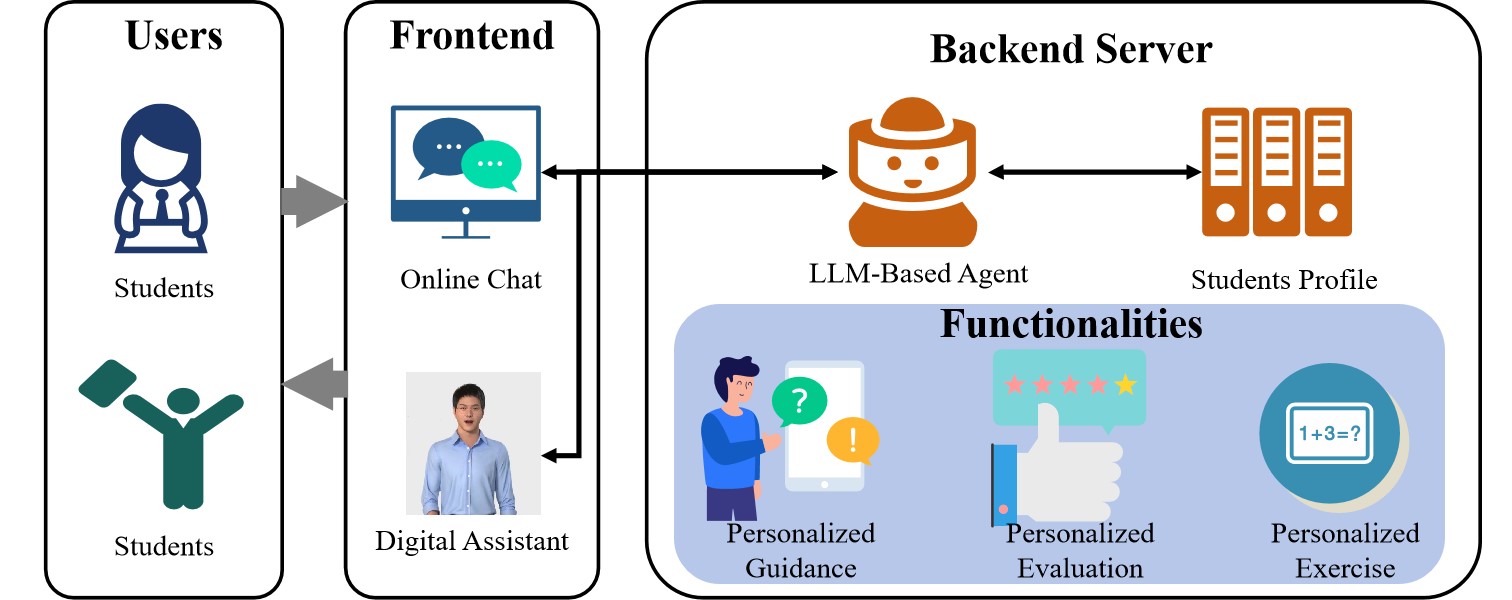}
    \vspace{-5pt} 
    \vspace{-5pt} 

    \caption{SortingHat Structures}
    \label{fig:sh}
\end{figure}

SortingHat serves as a versatile TA for university courses in Figure.\ref{fig:sh}. SortingHat comprises three main components: a digital chat and digital human interface, LLM-based agents, and a student database. Students interact with the system through a Dingding-based chat or an emotional digital human, which adapts its responses and tone to their learning progress and emotional state. At its core, LLM-based agents provide expert guidance, generate personalized exercises, and evaluate submissions with detailed feedback to help students improve. The student database stores non-privacy-sensitive information, such as learning progress and performance history, enabling tailored and adaptive learning experiences.
\vspace{-10pt} 
\vspace{-5pt} 

\subsection{Customized Guidance Assistant}
\label{agent}
\begin{figure}[h]
    \vspace{-10pt} 

    \centering
    \includegraphics[width=0.9\linewidth]{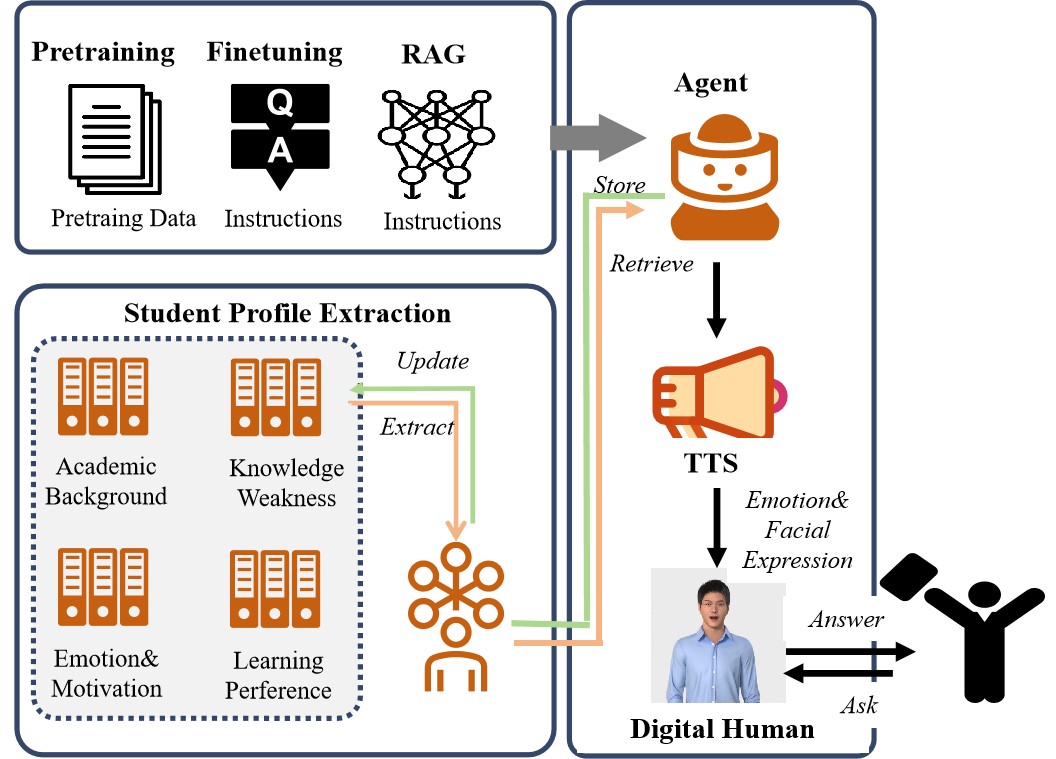}

    \caption{Customed Digital Assistant Generation}
    \label{fig:digitalassistant}

\end{figure}

Digital teaching assistants provide scalable, personalized support, adapting to student needs and fostering engagement. With 24/7 availability, they offer cost-effective and immersive learning experiences \cite{nass2000machines, nissen2022see}. As AI evolves, digital assistants hold the potential to redefine education with innovative, personalized learning experiences. Motivated by the potential of digital assistant, we developed a precise, personalized, and knowledgeable digital assistant designed to enhance OS education, as illustrated in Figure \ref{fig:digitalassistant}. 

    \vspace{-5pt} 

\vspace{-10pt} 

Current LLM-based AI assistants \cite{sajja2024artificial,imran2023analyzing} excel in understanding instructions and maintaining coherent reasoning but face notable limitations. Text-based interactions lack emotional engagement, reducing students' motivation, participation, and sense of belonging \cite{picard2000affective,graesser2003autotutor,nass2000machines}. Additionally, these systems are single-modal, lacking the immersive qualities proven effective in education \cite{dede2009immersive,bromley2013active}. They also struggle with domain-specific knowledge, and training on frequently updated, dynamic lecture materials is impractical due to the time and resources required.

The core of the guide is a knowledgeable LLM-based agent. To train a domain-specific model, we collected over 60,000 OS-related documents and generated 10,000 instructions. We also propose a retrieval-augmented generation (RAG) systems. based on the GraphRAG framework \cite{graphragmicro}, leveraging a corpus that includes OS source code and technical documentation on emerging OS architectures.

To address the limitations of traditional one-size-fits-all education methods, we designed a digital 3D human assistant that provides consistent and emotionally engaging interactions, moving beyond the constraints of text-based communication. This personalized digital assistant, built on \textit{Fay}, is integrated with a text-to-speech (TTS) server capable of converting text into emotional and contextually appropriate speech. By combining visual, auditory, and emotional elements, the digital human creates a multimodal learning experience that enhances engagement and understanding.

Behind the digital human, an agent extracts relevant student profiles from the database and analyzes key features, such as their learning progress and knowledge gaps. This information is then shared with another agent tasked with generating responses to the student’s queries. By incorporating this personalized data, the system creates a tailored conversational environment that adapts to the specific needs of each student. The agent dynamically adjusts the tone, depth of knowledge, and supplementary explanations required to clarify points, while ensuring the responses are delivered with appropriate emotional expression. Given the unique challenges of OSs, the system goes beyond conceptual guidance by providing practical examples and case studies to bridge theory and real-world application. This approach ensures not only accurate but also empathetic, context-aware, and practice-oriented interactions, fostering a more effective and engaging learning experience.

\subsection{Tailored OS Exercises Based on Student Characteristics}
\label{generatingframework}
\begin{figure}[h]
    \vspace{-5pt} 
    \centering
    \includegraphics[width=1\linewidth]{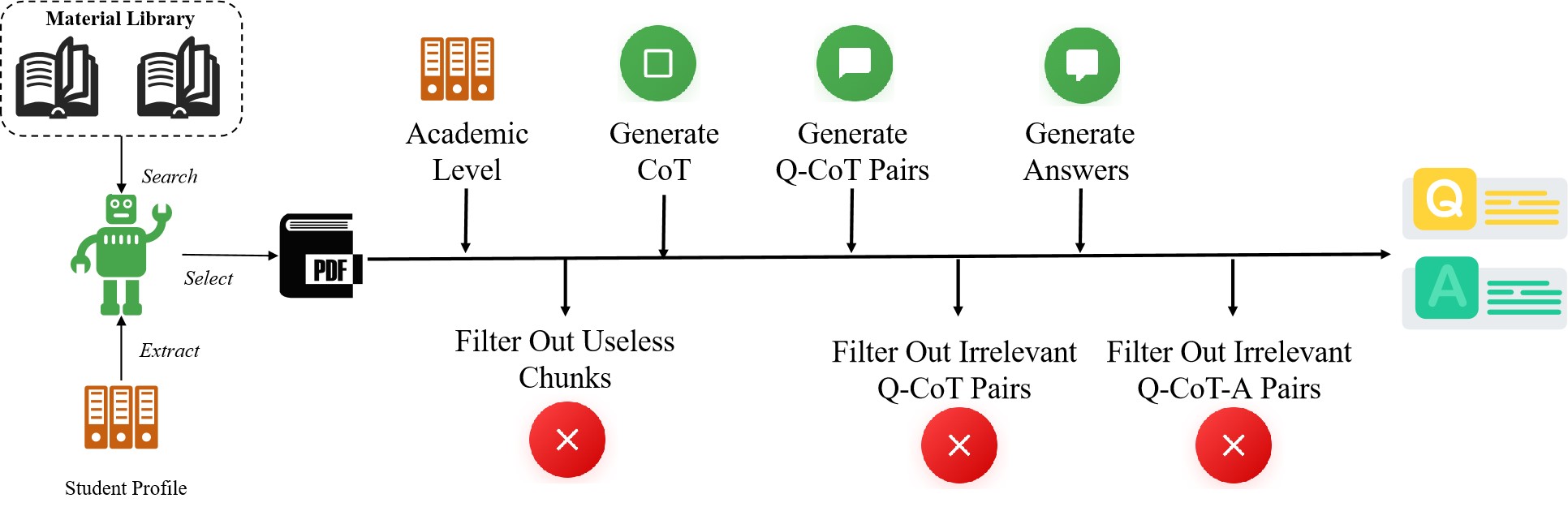}
    \caption{The multi-agent workflow for tailored exercise generation involves selecting suitable materials based on the student's profile and a multi-agent discussion workflow to create customized exercises.}
    \label{fig:generation}
    \vspace{-10pt}

\end{figure}

Traditional exercises in OS education often fail to meet the evolving demands of modern curricula and the rapid advancements in OS technologies. They are typically generic, focusing on basic concepts while lacking depth to assess understanding of cutting-edge systems. Additionally, traditional exercises do not account for individual students’ learning progress or weaknesses, highlighting the need for personalized, adaptive exercise generation tailored to each student’s unique background and needs. To address these issues, we propose an LLM-based multi-agent exercise generation workflow, as illustrated in Figure \ref{fig:generation}.

\vspace{-5pt} 

The system begins by analyzing the student’s profile to assess their academic level and learning needs. Based on this analysis, it selects appropriate learning materials tailored to the student’s specific requirements. These materials are then processed through a multi-agent framework, where generation agents create exercises aligned with the material and the student’s academic level. Evaluation agents critically review the exercises for accuracy, clarity, and relevance. Through iterative collaboration and refinement among the agents, the system ensures that the final exercises are coherent, precise, and educationally effective. This adaptive and scalable workflow provides a robust solution for creating personalized, high-quality educational resources.

\subsection{Personalized Submission Evaluation}
\label{ppl}
Evaluating student submissions is challenging due to their length, complexity, time-consuming nature, and the need for domain-specific analysis. Human evaluation often suffers from bias and inefficiency, particularly in large classes, where the volume of submissions can overwhelm evaluators. An LLM-as-Judge approach is promised to address these issues by offering scalability, fairness, and cost-effectiveness. However, existing models face limitations: \textbf{(C1)} lack of domain-specific knowledge, \textbf{(C2)} context-length constraints and attention decay \cite{vaswani2017attention}, and \textbf{(C3)} susceptibility to hallucinations and instability.

\begin{figure}[h]
    \vspace{-10pt} 

    \centering
    \includegraphics[width=1\linewidth]{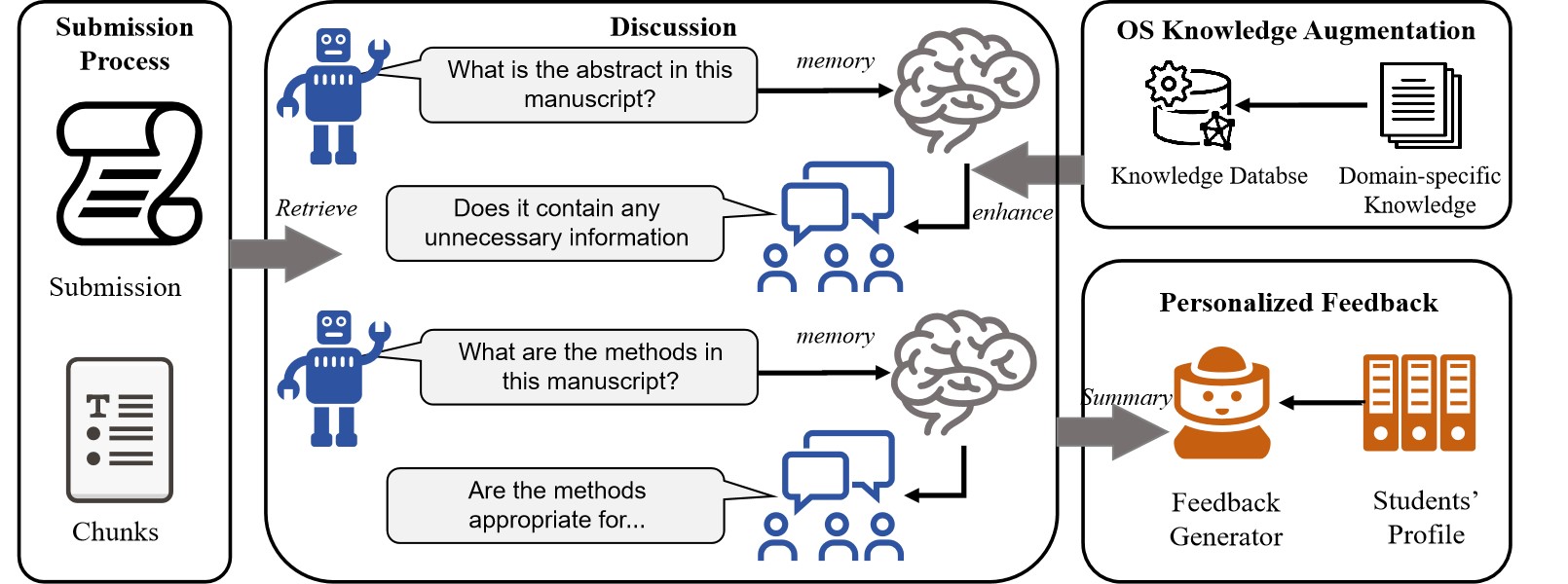}
    \vspace{-10pt} 
    \caption{Process of Evaluation: Long submissions are divided into smaller, manageable sections for efficient processing. Agents, augmented with additional knowledge resources, engage in collaborative discussions to ensure an accurate evaluation. The feedback generator integrates the student’s academic level and historical submissions to produce personalized and contextually relevant feedback.}
    \label{fig:evaluate}
\end{figure}

To tackle these challenges, we propose an RAG-enhanced MARL workflow (Figure \ref{fig:evaluate}) to deliver fair, scalable, and efficient evaluations while addressing the shortcomings of current models.
    \vspace{-15pt} 

To process lengthy submissions, the input is divided into smaller chunks and stored in a submission database. The agent workflow is divided into three key stages: Information Extraction (IE), Discussion Evaluation (DE), and Feedback Generation. IE agents extract essential information from sections such as the abstract, introduction, and methodology. This contextual data is stored in their memory module. Once these tasks are complete, IE agents participate in a collaborative discussion with evaluator agents. These agents, augmented with additional knowledge resources and identical to the models described in Section \ref{agent}, engage in a structured dialogue following an MARL framework. The evaluation process is broken down into 20 specific goals, each targeting different aspects of the submission. This collaborative approach addresses the limitations of a single LLM-based agent, such as hallucinations and instability. Finally, the feedback generator combines insights from the discussion with the student’s academic level and submission history to produce personalized and contextually relevant feedback. 
\section{Demonstration}
\begin{figure}[h]
      \vspace{-10pt} 
    \centering
    \includegraphics[width=0.9\linewidth]{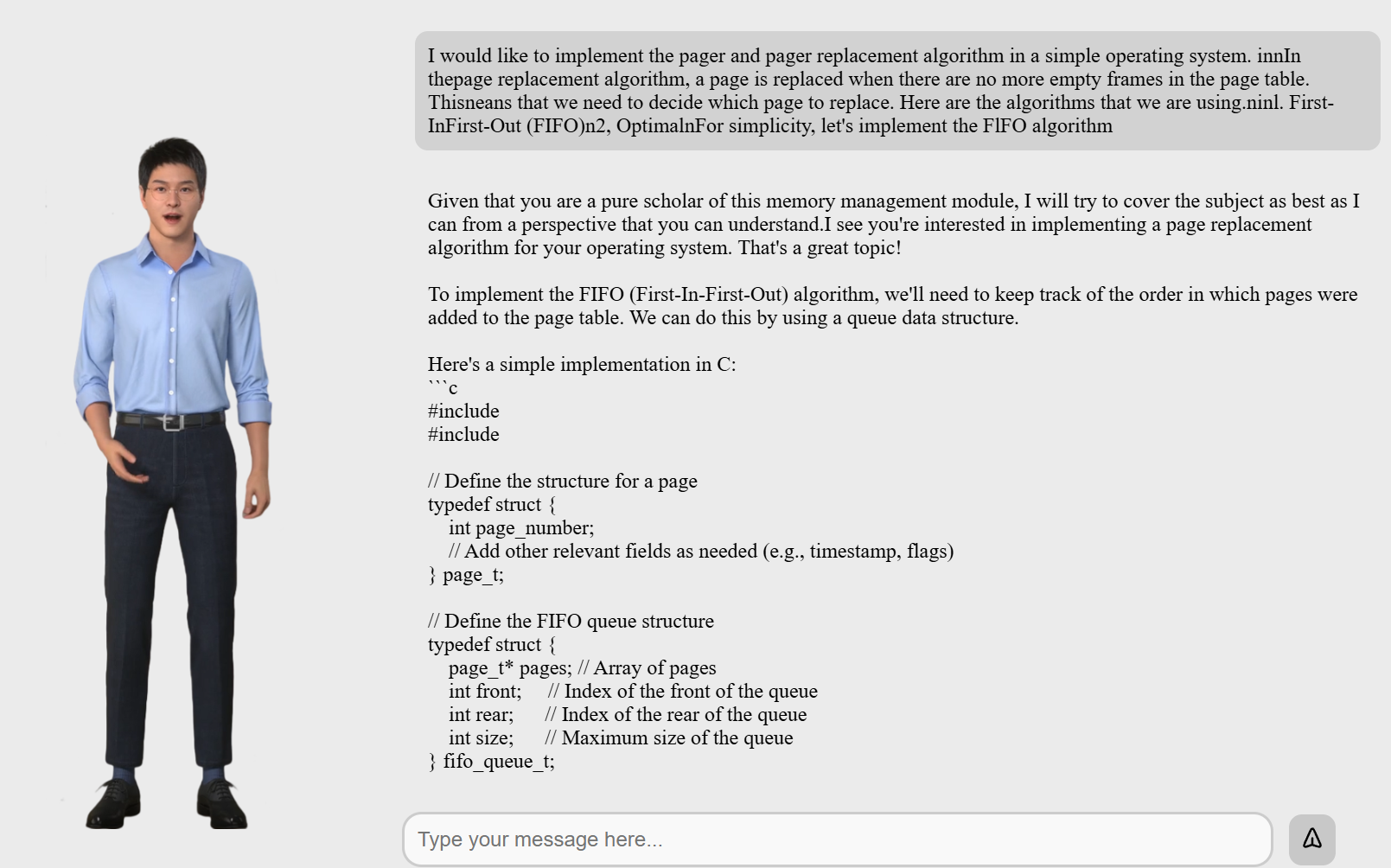}
    \caption{Digital Human TA offering personalized guidance while considering the academic level of students}
    \label{fig:interaction}
\end{figure}

As shown in Figure \ref{fig:interaction}, students engage with the digital teaching assistant via a multimodal, immersive interface. This assistant delivers responses that closely mimic classroom-style guidance and are more specifically tailored to the students' academic level, ensuring a personalized and effective learning experience.

\begin{figure}
    \centering
    \includegraphics[width=0.9\linewidth]{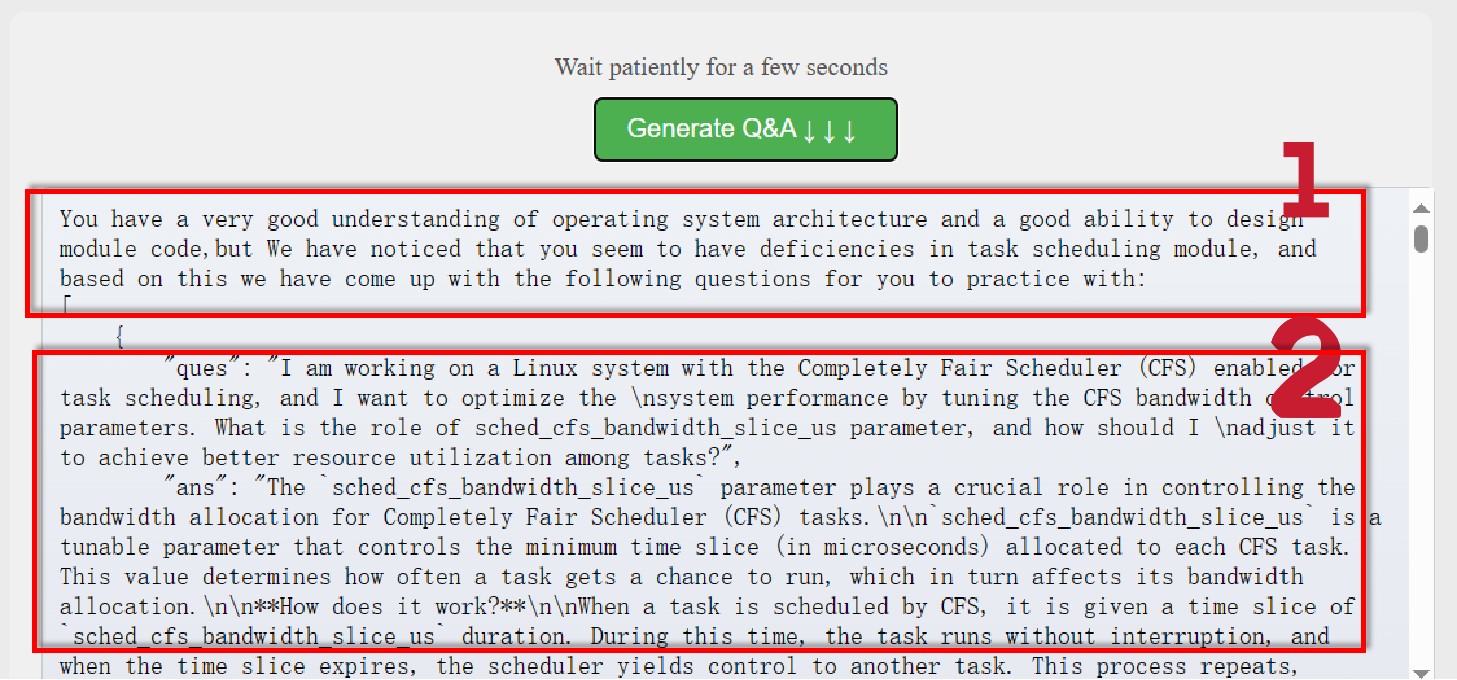}
    \caption{An Example of Exercise Generation based on Students Academic Level:\ding{172} concludes student's academic level while \ding{173} generate corresponding exercise.}
    \label{fig:exerciseexample}

\end{figure}

Additionally, the exercises generated by the framework described in Section \ref{generatingframework} demonstrate flexibility, depth, and a high degree of customization, as illustrated in Figure \ref{fig:exerciseexample}. This customization ensures that the exercises are tailored to the individual student’s academic level, learning history, and specific needs, providing a more personalized learning experience.

\begin{figure}[htbp]
      \vspace{-10pt} 

    \centering
    \includegraphics[width=0.8\linewidth]{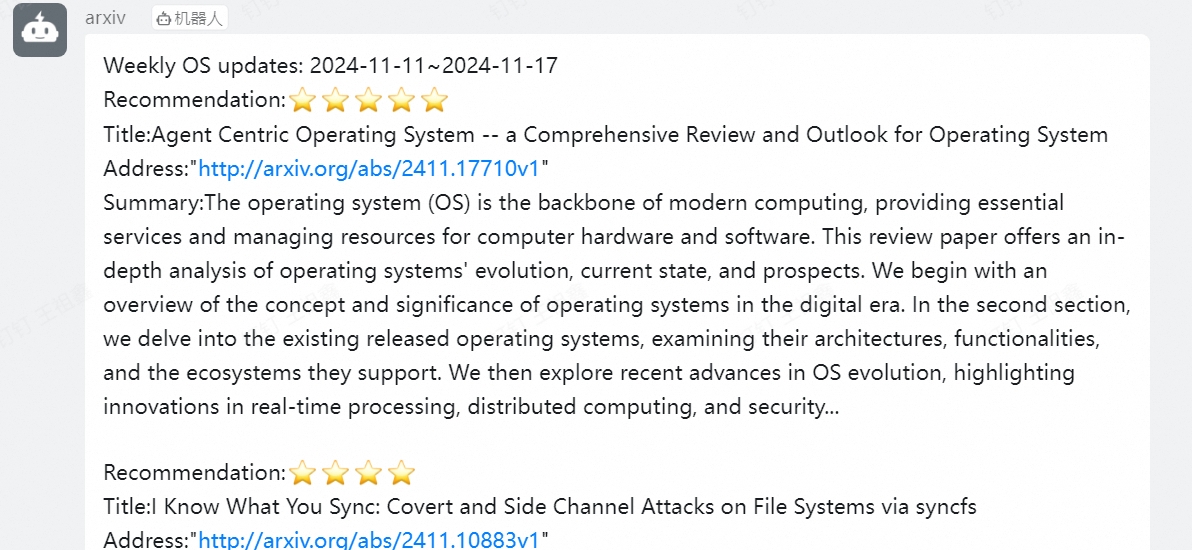}

    \caption{An Example of Weekly Posts about OS Advancement}
    \label{fig:dingding}
\end{figure}

\begin{figure}[htbp]
    \centering
    \includegraphics[width=0.8\linewidth]{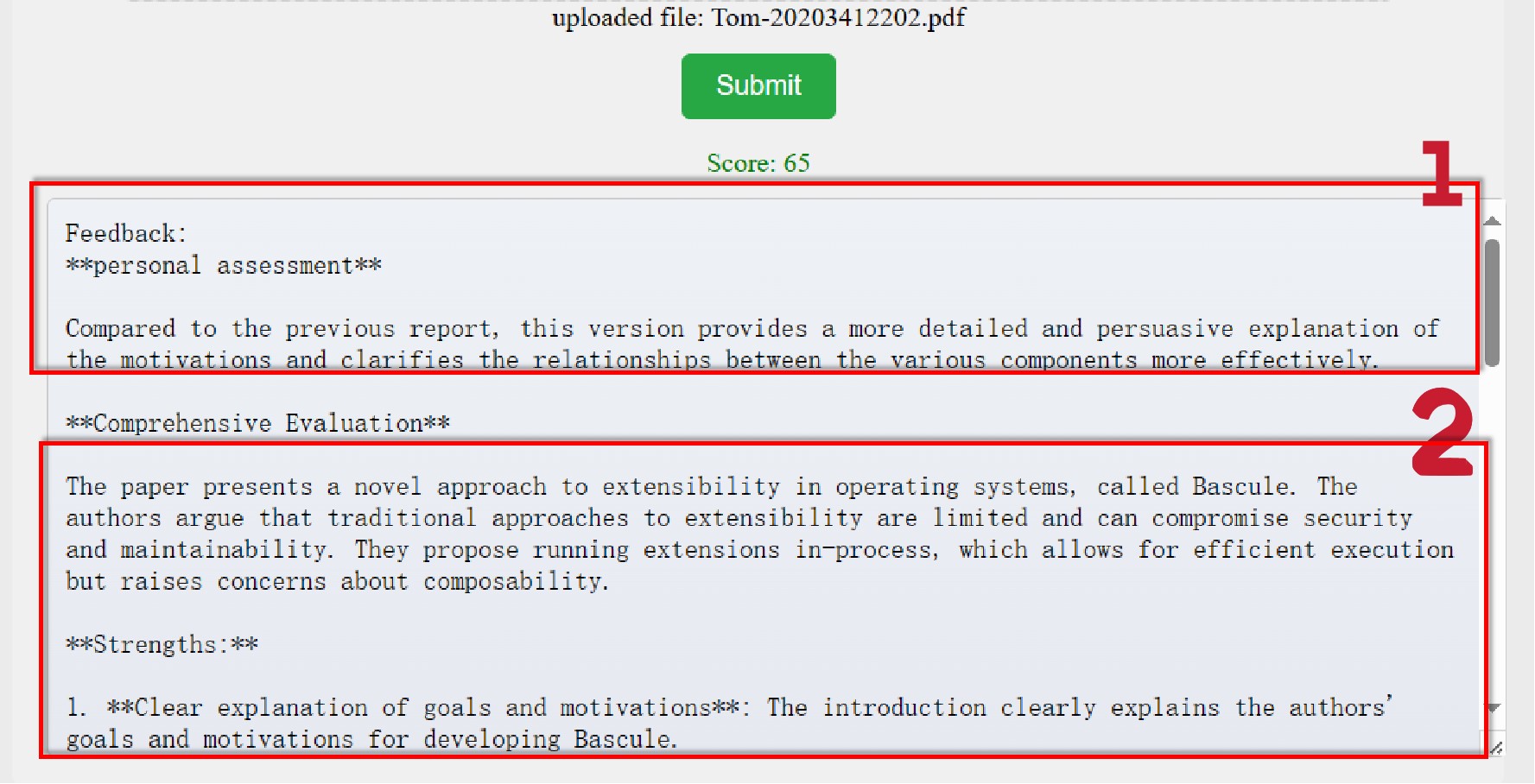}
    \caption{An Example of Evaluating a Submission:\ding{172} demonstrates how SortingHat provides personalized feedback tailored to the student's academic level and submission history, while \ding{173} highlights its ability to deliver a comprehensive evaluation, ensuring both depth and accuracy in the assessment.}
    \label{fig:evaluation}
\end{figure}

Our evaluation pipeline in Section \ref{ppl} can be used to evaluate students' submissions(Figure.\ref{fig:evaluation}). We would comprehensively evaluate them and give a score and its corresponding reasons. Also,in order to stay updated on the latest advancements in the OS field(Figure \ref{fig:dingding}), agents use Arxiv's convenient API,  crawl newly submitted papers on "Operating Systems" each week, evaluate them, and recommend the most relevant ones.

    \vspace{-15pt} 

\begin{figure}[htbp]
    \vspace{-15pt} 
    \centering
    \includegraphics[width=0.9\linewidth]{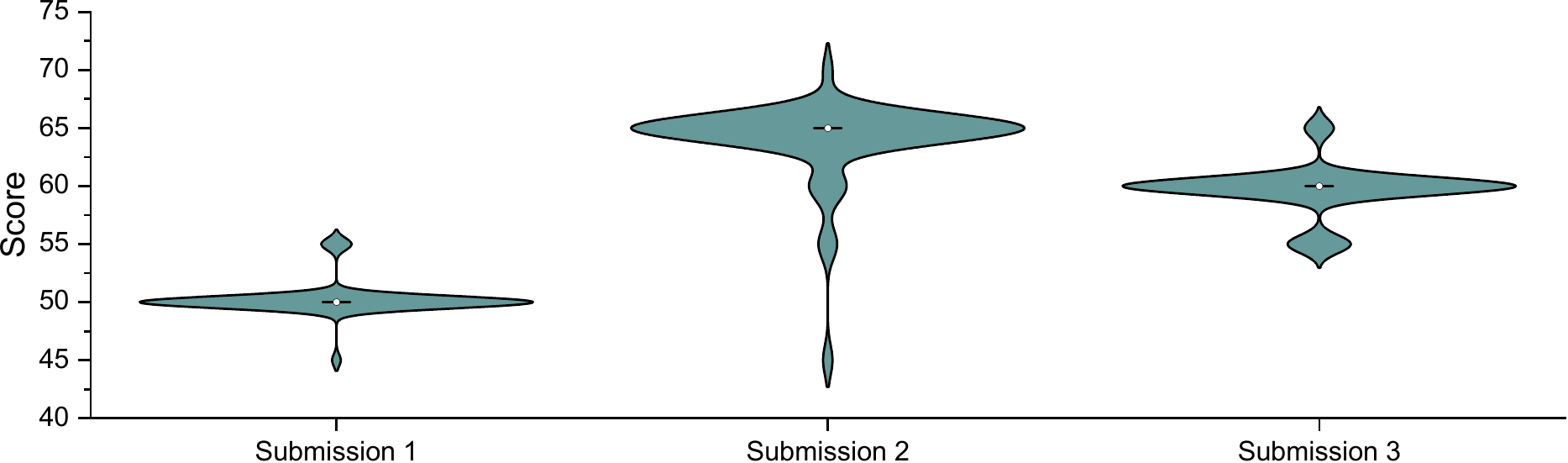}
    \caption{Distribution of Evaluation Scores for Submissions Evaluated 100 Times}
    \label{fig:compare}
\end{figure}

To demonstrate the stability of our framework, we evaluated three papers of varying quality 100 times each in Figure \ref{fig:compare}. The results show that our framework produces consistent evaluation scores, with variations reflecting the inherent quality differences.

\vspace{-5pt} 

\section{Conclusion}
This paper presented SortingHat, a personalized educational assistant designed to overcome the limitations of traditional OS education. By integrating a customizable digital human interface, LLM-based reasoning agents, and a student database, SortingHat delivers tailored learning experiences that adapt to individual students’ needs. It address challenges such as outdated exercises, lack of individual support, and inconsistent evaluation. SortingHat represents a step forward in personalized education, offering scalable solutions for OS curricula and a foundation for future advancements in adaptive learning technologies.

\section{Acknowledgments}
This work was supported in part by the fund of Laboratory for Advanced Computing and Intelligence Engineering, and in part by the National Science Foundation of China under Grants (62472375,62125206), and in part by the Major Program of National Natural Science Foundation of Zhejiang(LD24F020014, LD25F020002), and in part by the Zhejiang Pioneer (Jianbing) Project (2024C01032), and in part by the Ningbo Yongjiang Talent Programme(2023A-198-G).
\vspace{-5pt} 

\bibliographystyle{ACM-Reference-Format}
\bibliography{main}

\appendix

\end{document}